\documentclass{article}[12pt]      

\usepackage{amsmath,amsthm,amssymb,epsfig,color,calc,rotating,graphics,alltt}

\usepackage{hyperref}

\usepackage[normalem]{ulem}

\newcommand\backmatter{\appendix
\def\chaptermark##1{\markboth{%
\ifnum  \c@secnumdepth > \m@ne  \@chapapp\ \thechapter:  \fi  ##1}{%
\ifnum  \c@secnumdepth > \m@ne  \@chapapp\ \thechapter:  \fi  ##1}}%
\def\sectionmark##1{\relax}}

\newcommand{\Fox}[7]{H^{#1,#2}_{#3,#4}
\left( #5 \left| \begin{array}{cc}
#6\\
#7
\end{array}
\right) \right.
}

\begin{document}

\title{Beta Rank Function: A Smooth Double-Pareto-Like Distribution \\
\author{
Oscar Fontanelli$^1$, Pedro Miramontes$^{2}$, Ricardo Mansilla$^{1}$, Germinal Cocho$^{3\dagger}$, Wentian Li$^4$ \\
{\small  1. Centro de Investigaciones Interdisciplinarias en Ciencias y Humanidades}
{\small   Universidad Nacional Aut\'{o}noma de M\'{e}xico}\\
{\small  2. Departamento de Matem\'{a}ticas, Facultad de Ciencias, Universidad Nacional Aut\'{o}noma de M\'{e}xico} \\
{\small  3. Instituto de F\'{i}sica,}
{\small   Universidad Nacional Aut\'{o}noma de M\'{e}xico}\\
{\small  4. The Robert S. Boas Center for Genomics and Human Genetics,
The Feinstein Institutes for Medical Research}
}
\date{\today}      
}  
\maketitle                   
\markboth{\sl Li et al. }{\sl Li et al. }

\thanks{$\dagger$ Deceased 9 May 2019.}

\begin{center}
{\bf Abstract}
\end{center}

The Beta Rank Function (BRF), $x(u) =A (1-u)^b/u^a$, where $u \in (0,1]$
is the normalized and continuous rank of an observation $x$, has wide applications in fitting real-world data from social science to biological phenomena. The probability density function (pdf) converted from the BRF, $f_X(x)$, does not usually have an analytic expression except for specific parameter values.  We show however that it is approximately a unimodal skewed and asymmetric two-sided power law/double Pareto/log-Laplacian distribution. The pdf of the BRF has very simple properties when the independent variable is log-transformed: $f_{Z=\log(X)}(z)$ . At the peak it makes a smooth turn from one side to the other and it does not diverge, lacking the sharp angle observed in the double Pareto or the Laplace distribution. The mode of $f_Z(z)$ (peak position) is $z_0 =\log(A)+ (a-b) \log(\sqrt{a}+\sqrt{b}) - (a\log(a)-b\log(b))/2 $; the probability is partitioned by the peak to the proportion of $\sqrt{b}/(\sqrt{a}+\sqrt{b})$ (left) and $\sqrt{a}/(\sqrt{a}+\sqrt{b})$ (right); the functional form near the peak is largely controlled by the cubic term in the Taylor expansion when $a \ne b$; the mean of $f(z)$ is $E[z]=\log(A)+a-b$; the decay on left and right sides of the peak is approximately exponential with the form $e^{\frac{z-\log(A)}{b} }/b$ and  $e^{ -\frac{z-\log(A)}{a}}/a$.
These results are also confirmed by numerical simulations. As a comparison, properties of $f_X(x)$ without a log-transformation of the variable, some of them were also derived analytically, are much more complex, though the approximate power-law behavior, or double Pareto, $(x/A)^{1/b}/(bx)$ (for $x < A$) and $(x/A)^{-1/a}/(ax)$ (for $x > A$) is simple. Our results  elucidate the relationship between BRF and log-normal distributions when $a=b$, and explain why the BRF is ubiquitous and versatile. Based on the pdf, we also suggest a quick way to elucidate if  a real data set follows a one-sided power-law, a log-normal or a 
two-sided power-law of  BRF. We illustrate our results with a few examples: urban populations and returns of financial indexes.

\maketitle
\vspace{0.1in}

{\bf Abbreviations}: 
BRF: Beta rank function;
cdf: cumulative density function;
DGBD: Discrete Generalized Beta Distribution;
DPLN: double-Pareto-lognormal distribution;
pdf: probability density function;



\section{Introduction}

\indent

Since the introduction of the two-parameter Discrete Generalized Beta Distribution (DGBD) (or Beta-like Rank Function or Cocho Rank Function) \cite{mansilla,gustavo}, a wide range of real-life data have been successfully fitted by this function \cite{egghe,campanario,wli-entropy,petersen,roberto,wli-jql,wli-baby, wli-physica,wli-jql2,petersen13,ausloos14,ausloos14b,finley,wli-bmc,sole,morales,wli-gss,oscar-au,fenner, roberto18,nguyen,lopez,ghosh}. Two questions naturally arise: first, what's the corresponding probability density function (pdf) of the DGBD? Second, why is this family of functions so ubiquitous? In order to fully address these questions, we will need to define a continuous-rank version of the DGBD, which is a function with discrete independent variable (rank). 

These two questions are partially addressed in \cite{plos-lav}: when the two parameters in the DGBD are equal, leading to the one-parameter rank function called Lavalette Rank Function \cite{lav,popescu,popescu2,volo,lav07}, the corresponding pdf can be derived analytically \cite{chlebus,plos-lav}. This pdf, which we called Lavalette distribution, is similar (but not identical) to the lognormal distribution \cite{plos-lav}. But the nature of the difference between the Lavalette and lognormal distributions has not been addressed in depth.  It is also shown in \cite{plos-lav} that when one of the parameters in the DGBD is equal to zero, it reduces to a one-sided power-law distribution; when the other parameter is zero,  it becomes a uniform distribution.  Although there are other attempts to explain  the widespread applications of the DGBD \cite{naumis,beltran}, the spanning from uniform, approximately log-normal and one-sided power-law distribution by turning the parameters demonstrates the versatility of the family of functions defined by the DGBD.

In order to make a connection between DGBD and families of continuous probability distributions, we will introduce its continuous equivalent, which we will call Beta Rank Function (BRF). This coincides with a quantile function proposed in \cite{hankin2006new} (Hankin-Lee quantile function or Davies distribution), which accurately approximates some non-negative distributions such as lognormal, Weibull and generalized Tukey. This quantile function has also been proposed in \cite{gilchrist2000statistical}, where it is called the power-Pareto distribution.  Ideally, we would like to derive its corresponding pdf in an analytic form. However, the task is not straightforward except for a few specific points (or lines) in the parameter space. The apparent impossibility for a closed-form expression of the underlying distribution has been noticed in \cite{brzezinski2014empirical}. As mentioned above, these exceptions include $a=b$, $a=0$, and $b=0$. We will show later that it also includes $a=k\cdot b$, $b=k\cdot a$ where $k=2$ or sometimes other integer values.

In this paper we use a combination of analytic and simulation approaches to study the properties of pdfs derived from the BRF family. In particular, we will show that the pdf becomes much simpler when the independent variable is log-transformed. One of the most important results we obtained is that the fall off from the peak is exponential. In other words, the pdf is approximately log-Laplace, or double-Pareto, or double-power-law \cite{reed01,kotz}. Differing from the lognormal distribution, our pdf can be  asymmetric around the peak and the cubic term, as well as the quadratic form, plays a role in the functional form. Our pdf also differs from lognormal in its decay form from the peak. The exponential fall off on the two sides of the peak can be different, with the left side controlled by parameter $b$ and the right side by parameter $a$.

The results in this paper provide a comprehensive picture of the pdf associated to the BRF or Hankin-Lee-Davies function without the analytic expression of its functional form. This pdf is an approximate double Pareto/power-law function away from the peak,  a different perspective from the typical practice to focus attention on  one tail (e.g. in Zipf's law). The function near the peak is smooth, not only different from the potential divergence of the distribution in one-sided power-laws, but also  different from the divergence of the second derivative (sharp transition) for double-Pareto functions. The different decay rates on both sides of the peak also make our novel distribution flexible in data fittings. 

The paper is organized as follows: first we describe the rank order representation of a random variable and its connection with the pdf; in this context we will define the Beta Rank Function, which is the continuous equivalent of the DGBD with normalized rank. For the sake of completeness, we analyze some particular cases of the BRF, which we had already studied in previous works. Next we will show that our analysis is greatly simplified after a logarithmic transformation, we will introduce the novel log-BRF family of distributions and analyze some of its main properties. Then we will introduce the novel BRF family of distributions, study some of its main properties and propose methods to produce pseudo-random numbers from this distributions and to numerically approximate its pdf. After that, we will illustrate the possible usefulness of these distributions in data analysis by means of two particular examples (distribution of financial log-returns and city population distribution). The paper ends with Discussion and Conclusion sections.


\section{Rank-Ordering Statistics and the Beta Rank Function}

\subsection{Connection between rank-size function and probability density function}

\indent

Let $X$ be a positive and continuous random variable with a probability density function $f(x)$. Let $\underline{x}=(x_1,...,x_N)$ be a list of $N$ independent realizations of $X$ (a random sample). We define the rank $r$ of the observation $x_i$ within the list $\underline{x}$ as the number of observations in the list that are greater or equal than $x_i$. For instance, the rank of the largest observation in the sample will be $r(\mbox{max}(\underline{x}))=1$, whereas for the lowest observation $r(\mbox{min}(\underline{x}))=N$. Ranks are well defined if the probability of making two identical observations is equal to zero, which is here the case, since we assume the observations come from a density function.  Consider now the list $\underline{x}$ sorted in decreasing order, $\underline{x_s}=(x_{[1]},...,x_{[N]})$, where $x_{[1]} \geq x_{[2]} \geq ... \geq x_{[N]}$ (observe that the k-th order statistic is $Y_k = X_{[N-k+1]}$). We call the list of ranks of the ordered list $\underline{x_s}$ the rank list, $\underline{r}=(r_1,...,r_N)$. A plot of $\underline{x_s}$ against $\underline{r}$ is called a \emph{rank-size plot} of the observations. By construction, these plots are always decreasing.

Now consider the rank list normalized to the $(0,1]$ interval,
$\left(\frac{r_1-r_1}{r_{N}-r_1},...,\frac{r_N-r_1}{r_N-r_1}\right)$.
Unless specified we will consider only normalized ranks, so we will refer to them simply as the rank (to be made continuous) variable $u$.  A rank-size function is a function that quantifies the dependence of the size (or value) of an observation $x$ as a function of the rank $u$ it would have within an infinite list of observations. The rank-size function $x=x(u)$ can be constructed from the pdf $f(x)$ in the following way: notice that, according to our definition, the (normalized and continuous) rank of $x$ equals the probability of making a larger observation,
\begin{equation}
\label{rank-size}
\displaystyle u(x)=1-F(x)=\int_x^\infty f(t)dt,
\end{equation}
where $F(x)$ is the cumulative distribution (cdf). Note that, after this construction, 
the $u(x)$ function coincides with the well-known survival function. If the cdf is strictly increasing and continuous, then we can solve Eq.(\ref{rank-size}) for the pdf,
\begin{equation}
\label{density}
f(x)=-\frac{du}{dx}=-\frac{1}{\frac{dx}{du}}.
\end{equation}

According to Eq.(\ref{rank-size}), we define the \emph{rank-size function} $x(u)$ as the  inverse survival function, which can be computed if the pdf is known and the survival function is \-in\-ver\-ti\-ble.  On the other hand, Eq.(\ref{density}) allows us to compute the pdf of a random variable from its inverse survival function, or rank-size function.

\begin{figure}[ht]
\begin{center}
    \includegraphics[width=0.5\linewidth]{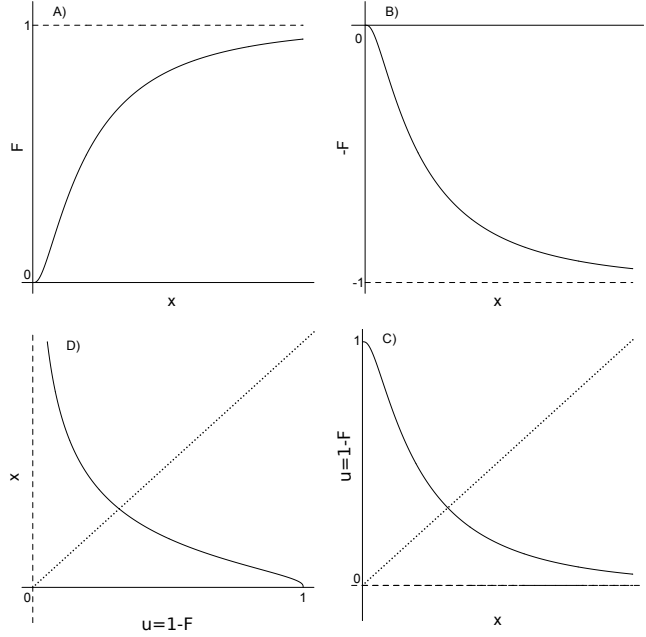}
\end{center}
\caption{
\label{fig1}
The rank-size plot approximates the cumulative function in a similar way in which the histogram approximates the density function. A) Start with a cdf $F$, B) reflect with through the $x$ axis ($-F$), C) make a $+1$ shift ($1-F$) and D) reflect through the identity line to get the rank-size function $x(u)$.
}
\end{figure}

Geometrically, it is possible to get the rank-size function from the cdf or vice versa in the following way:  starting from the cumulative function $F(x)$, reflect it respect to the $x$ axis (getting $-F(x)$), make a $+1$ shift in the $y$ direction (getting $u=u(x)=1-F(x)$) and reflect it respect to the $y=x$ line (getting $x=x(u)$).  Therefore, the rank-size plot of a collection of random observations serves as an attempt to approximate the cdf in a similar way that a histogram approximates the density function. This geometric procedure is illustrated in Fig.\ref{fig1}. A similar illustration can also be found in \cite{wli-entropy}.

Let's summarize the contents in this subsection as they will be repeatedly used
later: $x(u)$ is the rank-size function; $u(x) = 1- F(x)$ is the survival function and $f(x)=-du(x)/dx$ is the probability density function. Later on, we will also show that after $X$ is log-transformed, $Z=\log(X)$,
the probability density function of $Z$ is $f_Z(z)= xf_X(x)$.

In the next paragraph we will recall the definition of the Discrete Generalized Beta Distribution and define the Beta Rank Function, which will be a normalized and continuous-rank version of the DGBD, whose underlying pdf we intend to investigate. 

\subsection{Definition of the Beta Rank Function}

\indent

The Discrete Generalized Beta Distribution (DGBD) is a discrete, two-parameter rank-size function defined by
\begin{equation*}
x(r)=A\frac{(N+1-r)^b}{r^a},
\end{equation*}
where $x$ is the size (value) of the observation, $N$ is the total number of observations, $a$ and $b$ are parameters, the rank spans $r=1,2,...,N$, and $A$ is a scale normalization constant.

In order to establish a connection between this representation and the pdf, we need to define the continuous and normalized equivalent of the DGBD, which leads to the following definition.  The \emph{Beta Rank Function (BRF)} is a two-parameter rank-size function (inverse survival function) defined by
\begin{equation}
\label{BRF}
x(u) = A \frac{(1-u)^b}{u^a},
\end{equation}
where $u\in(0,1]$ and $a, b\geq 0$  are free parameters. The parameter $A$ is related to the scale of the data: it can be estimated from the data or, if the data is re-scaled,  can be simply set to 1. It is straightforward to see that $u_2 > u_1$ implies $x(u_1) > x(u_2)$, so the BRF is strictly decreasing over $(0,1]$. Therefore, the inverse function
$u(x)$ exists within this interval. As we mentioned, the BRF coincides with the Hankin-Lee-Davies quantile function proposed in \cite{hankin2006new} or power-Pareto distribution introduced in \cite{gilchrist2000statistical}. The connection between BRF and the Hankin-Lee-Davies function has already been noticed in \cite{sarabia2012modeling}.


\section{Particular cases of the BRF}

\indent

In this section we review some of the known results concerning special cases of the Beta Rank Function  in Eq.(\ref{BRF}).

\subsection{BRF yields a constant random variable when $a=b=0$}

\indent

When $a=b=0$, $x(u)$ is a constant function,
meaning there is one single outcome with non-zero probability. In other words, the BRF corresponds to a degenerate random variable with a delta distribution at the position $x=A$.


\subsection{BRF yields a uniform distribution when $a=0$ and $b=1$}
\label{BRF-uniform}
\indent

If $a=0$ and $b>0$, Eq.(\ref{BRF}) is reduced to
$$
 x =  A (1-u)^b 
$$
This equation can be solved exactly for $u$, $u=1-\left(\frac{x}{A}\right)^{1/b}$,
yielding the cdf $F(x)= \left(\frac{x}{A}\right)^{1/b}$.  Differentiating we get the pdf 
\begin{equation}
\label{uniform}
f_{a=0}(x) = \frac{1}{bA^{\frac{1}{b}}} x^{\frac{1}{b}-1}  \qquad \mbox{for} \qquad 0\leq x\leq A
\end{equation}
Notice that $x>A$ implies $F(x)=\left(\frac{x}{A}\right)^{\frac{1}{b}} >1$, 
hence the restriction $x\leq A$. Direct integration shows that this random variable 
has expectation $E[X]=\frac{A}{b+1}$  and its variance is given by 
$Var[X] = \frac{A^2b^2}{(b+1)^2(2b+1)}$.  For this case, we can give a 
closed-form formula for the characteristic function,

$$
E_{a=0}[e^{itx}] = \frac{1}{b}(-iAt)^{-\frac{1}{b}}\left[ \Gamma \left( \frac{1}{b}\right) - \Gamma \left(\frac{1}{b},-iAt\right) \right].
$$

Notice that Eq.(\ref{uniform}) is the density of a uniform random variable over $[0,A]$ if $b=1$. Also notice that the result in the previous subsection ($a=0, b=0$) can be re-created:
$$
f_{a=b=0}=
\lim_{b\rightarrow 0^+}f_{a=0}(x) = 
\left\{
\begin{array}{ccc}
0 & \mbox{if} & 0\leq x  < A\\
 & & \\
+\infty & \mbox{if} & x = A \\
\end{array}
\right.
$$


\subsection{BRF yields a Pareto distribution when $b=0$}
\label{BRF-Pareto}
\indent

If $b=0$ and $a>0$, then the BRF reduces to
$$
x= \frac{A}{u^a}
$$
which can be exactly solved for $u$, 
$u=\left(\frac{A}{x}\right)^{\frac{1}{a}}$.
Differentiating it leads to pdf
$$
f_{b=0}(x)=\frac{A^{\frac{1}{a}}}{a} \frac{1}{x^{{\frac{1}{a}}+1}} 
$$
This is the (one-tail)  Pareto distribution with shape parameter $a$ and scale parameter $A$.
If $a<1$, the mean exists and it is equal to $E[X]=\frac{A}{1-a}$; 
if $a<\frac{1}{2}$, the variance exists and it is equal to
$Var[X] = \frac{A^2a^2}{(1-a)^2(1-2a)}$. 
The characteristic function  is given by the formula
$$
E_{b=0}[e^{it}] = \frac{1}{a} EI_{1+\frac{1}{a}}(-iAt),
$$
where $EI_{\nu}(z)$ is the exponential integral function.


\subsection{BRF yields a Lavalette distribution when $a=b$}

\indent

As discussed with detail in \cite{plos-lav}, $u(x)$ can be solved analytically when $a=b$: 
$u=\frac{1}{1+ (x/A)^{1/a}}$,
and its negative derivative leads to the pdf
\begin{equation}
\label{eq-lav}
f_{a=b}(x) = \frac{ \left( \frac{x}{A}\right)^{1/a}}{ ax \left( 1+ \left(\frac{x}{A}\right)^{1/a}\right)^2}
\end{equation}
Detailed information about its expectation and variance, as well as applications, can be consulted in the reference cited above.

Before analyzing the properties of the pdf associated to the BRF, we will study the pdf of the random variable $Z= \log X$, where $X$ follows a BRF. 


\section{The log-BRF family of distributions}

\subsection{Logarithmic transformation of the independent variable \-
is a key step to simplify the probability density function for the BRF}

\indent

Following the results from the previous section, we would like to derive the general form of the pdf \-co\-rres\-pon\-ding to the BRF in Eq.(\ref{BRF}). However, we cannot hope to invert the rank-size function in order to write a general formula for this pdf in a closed-form.  To see this, note that in order to obtain the pdf from the rank function we need to obtain the cdf first, which is the solution for $\frac{x}{A} (1-F(x))^a - F(x)^b=0$.  From Abel-Ruffini's impossibility theorem, polynomial equations higher than 4th order  do not have a general, explicit  algebraic solution (e.g. \cite{zoladek}). However, we obtained multiple hints that log-transforming $X$ may simplify the characterization of the probability density function.

The first hint came from the Lavalette function, which is similar to the log-normal distribution \cite{plos-lav}. Both from the analytic expression and by numerical validation, it is also known that the Lavalette is not equivalent to the lognormal distribution \cite{plos-lav}. We can derive the pdf for $Z=\log X$ from Eq.(\ref{BRF}).
Note that $f(z)dz=f(x)dx$ implies $f(z)= f(x) dx/dz = x f(x)$. We denote $A=e^{z_0}$ and define $z'= z-z_0$, then
\begin{equation}
\label{eq-lav-z}
f_{a=b}(z)= x f_{a=b}(x)= \frac{ \left( \frac{e^z}{e^{z_0}} \right)^{1/a} }{a\left( 1+ \left( \frac{e^z}{e^{z_0}} \right)^{1/a} \right)^2}
= \frac{ e^{z'/a}}{a\left( 1+ e^{z'/a} \right)^2}
= \frac{ 1 }{a\left( e^{-z'/(2a)}+ e^{z'/(2a)} \right)^2}.
\end{equation}
This expression is already much simpler than Eq.(\ref{eq-lav}): it is the reciprocal square of the catenary curve $f(z)=c(e^{z/c}+e^{-z/c})/2$ \cite{huang}.

The expression in Eq.(\ref{eq-lav-z}) makes the analysis of the Lavalette distribution easier. It is more obvious from this functional form that the distribution of the logarithm of a Lavalette random variable, just as the normal  distribution, is symmetric around the peak in $z$. Near the peak, Lavalette and log-normal distributions are similar because the linear term in the Taylor expansion is zero, leaving only the quadratic term:
\begin{eqnarray*}
f_{a=b}(z) &\approx & 
\frac{1}{a} \frac{1}{ (1- \frac{z'}{2a} + \frac{z'^2}{8a^2} + \cdots + 1 + \frac{z'}{2a} + \frac{z'^2}{8a^2} + \cdots)^2}
 \nonumber \\
&=& \frac{1}{2a} \frac{1}{ (1 + \frac{z'^2}{8a^2} +   \cdots)^2} \nonumber \\
& \propto &  1-\frac{z'^2}{4a^2}.
\end{eqnarray*}
When this is compared to a similar Taylor expansion of a standard normal density near the peak, $1- \frac{z'^2}{2}$, we have $a = 1/\sqrt{2}= 0.707$.

\begin{figure}[ht]
\begin{center}
   \includegraphics[width=0.7\linewidth]{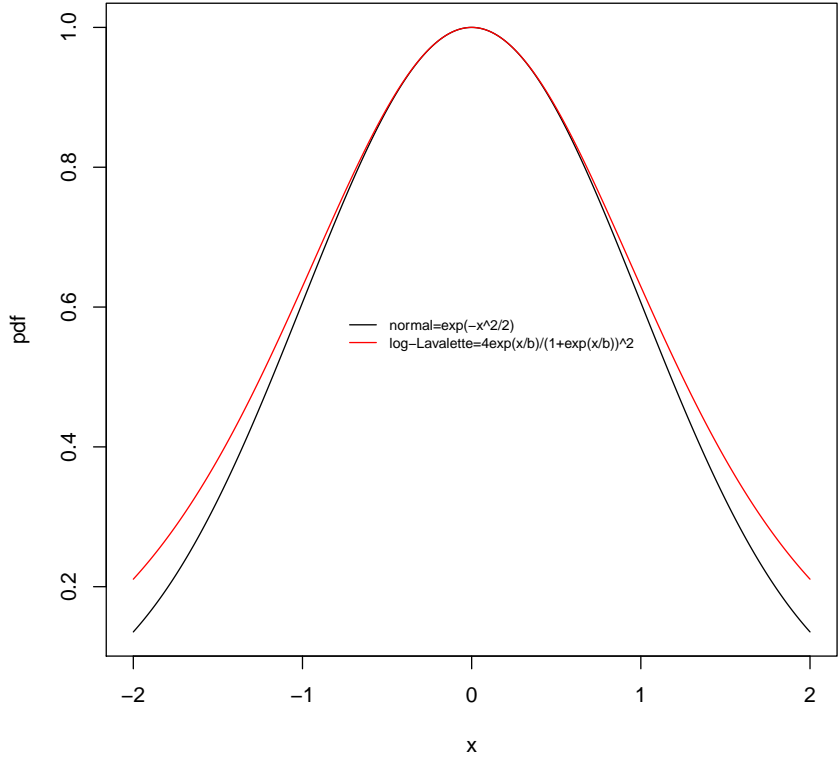}
\end{center}
\caption{
\label{fig2}
Comparison of the normal distribution (black) and log-transformed Lavalette distribution (red).
}
\end{figure}

Fig.\ref{fig2} compares the pdfs of the standard normal distribution $e^{-z^2/2}$
and the log-transformed Lavalette distribution $4\frac{e^{z/a}}{\left(1+e^{z/a}\right)^2}$
for $a=1/\sqrt{2}$. Since the parameter $a$ is chosen to fit the two functions near the peak, the two indeed match very well. When the independent variable is far away from the peak, however, the log-transformed Lavalette decays not as fast  as the normal distribution. The larger deviation between the two when the independent variable is away from the peak, despite the similarity near peak position, has already been observed in \cite{plos-lav}. 

\subsection{Definition of the log-BRF distribution}

Informally, consider a continuous random variable $X$ such that its inverse survival function $x(u)$ (the rank-size function) is given by
$$
x(u)=A\frac{(1-u)^b}{u^a}.
$$
Recall that the inverse survival function at $u$ is equivalent to the $(1-u)th$ quantile of the distribution. Consider the continuous random variable $Z=\log X$. We will see that we can compute the characteristic function of $Z$, which will provide us a way to give a formal definition of its distribution. \\

\noindent Let $f_X$ and $f_Z$ be the densities of $X$ and $Z$. Because $f_X(x)dx=f_Z(z)dz$ and $f_X(x)=-\frac{du}{dx}$, then $-f_Z(z)dz=du$, therefore

$$
\begin{array}{llclc}
E_Z\left[e^{itz}\right] & = & \displaystyle \int_{-\infty}^\infty e^{itz}f_Z(z)dz & = & \displaystyle \int_0^1 e^{itz(u)}du\\
 & & & & \\
 & = & \displaystyle \int_0^1 e^{it[\log A +b \log (1-u) -a\log u]}du & = & \displaystyle \int_0^1 e^{\log A^{it}} e^{\log (1-u)^{ibt}} e^{\log u^{-iat}}du\\
 & & & & \\
 & = & \displaystyle  A^{it} \int_0^1 (1-u)^{ibt}u^{-iat}du & = & A^{it} B(1-iat,1+ibt).\\
\end{array}
$$

Here we have used the definition of the complex exponentiation $a^b = e^{b\log a}$ and we are taking the principal branch of the logarithm.

We define that the continuous random variable $Z$ follows a \emph{log-BRF distribution} with parameters $A > 0$, and $a,b \geq  0$ if its characteristic function is represented by the formula
$$
\psi_Z(t) = A^{it} B(1-iat,1+ibt).
$$  

 It is possible to give a closed-form expression of the pdf of $Z$ in terms of the Fox H-function,

\begin{equation*}
f_Z(Z) = \displaystyle \Fox{1}{1}{1}{2}{\frac{e^z}{A}}{(0,a)}{(1,b)\,(-1,a-b)}.
\end{equation*}

\noindent See Appendix \ref{Fox-section} for a detailed derivation of this density and for the definition and properties of the $H$ function. 

\subsection{Mean, variance, and median of $f_Z(z)$ }

\indent

From the log-transformed BRF we can directly compute the first two moments of $Z$. We have
\begin{eqnarray*}
E[Z] &=& \int_{-\infty}^\infty zf_Z(z)dz = \int_{0}^\infty z f_X(x)dx
 \nonumber \\
&=& \int_{0}^{1}(\log(A)+b\log(1-u)-a\log(u))  \frac{du}{dx} dx
 \nonumber \\
&=& \log(A) + b \int_{0}^{1} \log(1-u)du - a \int_{0}^{1} \log(u) du
 \nonumber \\
&=& \log (A) - b +a
\end{eqnarray*}
and
\begin{eqnarray*}
E[Z^2] &=& \int_{0}^{1}(\log(A)+b\log(1-u)-a\log(u))^2  du 
 \nonumber \\
 &=& \int_0^1 \left( \log ^2 A+2b\log A \log(1-u) +b^2 \log(1-u)^2 -2a\log A \log u \right.
 \nonumber \\
 & & \left. -2ab\log(1-u)\log u +a^2 \log ^2 u \right)  du
  \nonumber \\
  &=& \log ^2 A -2b\log A -2b^2 +2a\log A -2ab\left(2-\frac{\pi^2}{6}\right) + 2a^2
 \nonumber \\
&=& ( \log(A))^2 + 2\log(A)( a-b )  + 2 (a-b)^2 + \frac{ab \pi^2}{3} 
\end{eqnarray*}
therefore,
\begin{equation*}
Var[Z]= E[Z^2]- E[Z]^2= (a-b)^2 + \frac{\pi^2 ab}{3}.
\end{equation*}
Interestingly, the variance does not depend on $A$ and it increases with a product of $a$ and $b$
(assuming $a$ is similar to $b$).

The median of $f_Z(z)$ corresponds to the $z$ value where $F=u=1/2$:
\begin{equation*}
Median[z]= \log(A)+ b\log(1/2)-a\log(1/2)= \log(A)-\log(2)(b-a) \approx \log(A) + 0.693 (a-b)
\end{equation*}
From these formula, it is clear that $A$ determine the location of the peak,
but not the shape, of the distribution.


\subsection{The probability partition by the peak and peak position (mode) in $f_Z(z)$ }

\indent

We ask the question, for $f_Z(z)$, what is the $z$ value where $f_Z(z)$ is the largest? 
And how is the probability being divided by the peak? In fact, the answer of the second question 
provides an answer of the first question.

At the peak position, the first derivative of $f_Z(z)$ is zero. Using an expression for $f_Z$ in terms of $u$ (that we will derive in Section \ref{section-exponential-decay},  Eq.(\ref{eq-fz-u})), we see that
\begin{equation*}
\frac{df_Z(z)}{dz}=0= -\frac{1}{(\frac{b}{1-u}+\frac{a}{u})^2} \left( \frac{b}{(1-u)^2} - \frac{a}{u^2}\right) \frac{du}{dz}
\end{equation*}
which leads to
$$
\frac{b}{(1-u)^2}= \frac{a}{u^2}
$$
and the solution is
\begin{equation*}
u_0= \frac{\sqrt{a}}{ \sqrt{a}+\sqrt{b}}, \hspace{0.1in}
F_0 \equiv 1-u_0 = \frac{\sqrt{b}}{ \sqrt{a}+\sqrt{b}}.
\end{equation*}
$F_0$ is the cdf accumulated up to the peak position and $u_0$ is the total probability on the right side of the peak. This result shows that the peak partitions the probability in $f_Z(z)$ by the $\sqrt{b}:\sqrt{a}$ ratio. 

Since $z$ has a simple relation with $u$ (which we will derive in section \ref{section-exponential-decay}, Eq.(\ref{eq-z-u})), we can obtain the peak position in $f_Z(z)$:
\begin{equation}
\label{eq-peak-z}
\begin{array}{lll}
z_0 & = & \log(A) +b \log \frac{\sqrt{b}}{\sqrt{a}+\sqrt{b}} -a \log \frac{\sqrt{a}}{\sqrt{a}+\sqrt{b}}\\
 & & \\
& = &  \log(A) + \frac{b\log(b)-a\log(a)}{2} + (a-b)\log( \sqrt{a}+\sqrt{b}).
\end{array}
\end{equation}
The second and the third terms above tend to cancel each other, so often $z_0$ is close to $\log(A)$.

All the statistical properties of $f_Z(z)$  that we have computed are summarized in Table 1.

\begin{table}[h]
\begin{center}
\begin{tabular}{|c|c|c|}
\hline
\multicolumn{2}{|c|}{properties of $f_Z(z)$}\\
\hline
mean & $\log(A)+a- b $   \\
median & $\log(A) +log(2)(a-b)$ \\
mode & $\log(A)  + (a-b)\log(\sqrt{a}+\sqrt{b}) - \frac{a\log(a)-b\log(b)}{2} $  \\
variance & $(a-b)^2+\frac{\pi^2 ab}{3}$ \\
\hline
\end{tabular}
\end{center}
\caption{Statistical properties of the log-BRF distribution defined by $Z=\log X$ where $X$ follows a BRF.}
\end{table}

\subsection{$f_Z(z)$ decays exponentially away from the peak}
\label{section-exponential-decay}

We first illustrate our conclusion by three special cases (uniform, Pareto and Lavalette), then we corroborate it by numerical simulation and finally we prove it analytically.

Recall that the identity $f_X(x)dx=f_Z(z)dz$ implies that the pdf in $z=\log(x)$ is $f_Z(z)= x f_X(x) = e^z f_X(x)$. For a uniform distribution, where $f_X(x)=C$, we have $f_Z(z)= Ce^z$. For the Pareto distribution the pdf is $f_X(x)=C/x^{\frac{1}{a}+1}$, therefore $f_Z(z)=Ce^{-z/a}$. In both situations the decay from the peak ($z=0$) is exponential. For the uniform distribution the decay is towards the negative $z$ values, whereas for the Pareto distribution it is towards the positive $z$ axes. The transition between a power-law and an exponential distribution through a log-transformation (or exponential-transformation in the other direction) of the variable has been noticed before \cite{wli-ieee,wli-comp,cze} (see section 14.2 of \cite{sornette} for other possible transformations).

The Lavalette distribution is somewhat different from the uniform and Pareto distribution
because it decays on both sides of the peak, and the decay near the peak is normal-like
($e^{-z^2}$). However, when $z'$ is large, from Eq.(\ref{eq-lav-z}):
\begin{eqnarray*} 
f_{a=b}(z) &\approx&  \frac{1}{a e^{z'/a}} =\frac{e^{- (z-\log(A))/a}}{a}
=\frac{A^{1/a}}{a}e^{-\frac{z}{a}}
 \quad \mbox{if} \quad z' \gg 0, \nonumber \\
f_{a=b}(z) &\approx& \frac{1}{a e^{-z'/a}} = \frac{e^{(z-\log(A))/a}}{a}
=\frac{A^{-1/a}}{a}e^{\frac{z}{a}}
 \quad \mbox{if} \quad z' \ll 0.
\end{eqnarray*} 

\begin{figure}[tp]
\begin{center}
    \includegraphics[width=12cm]{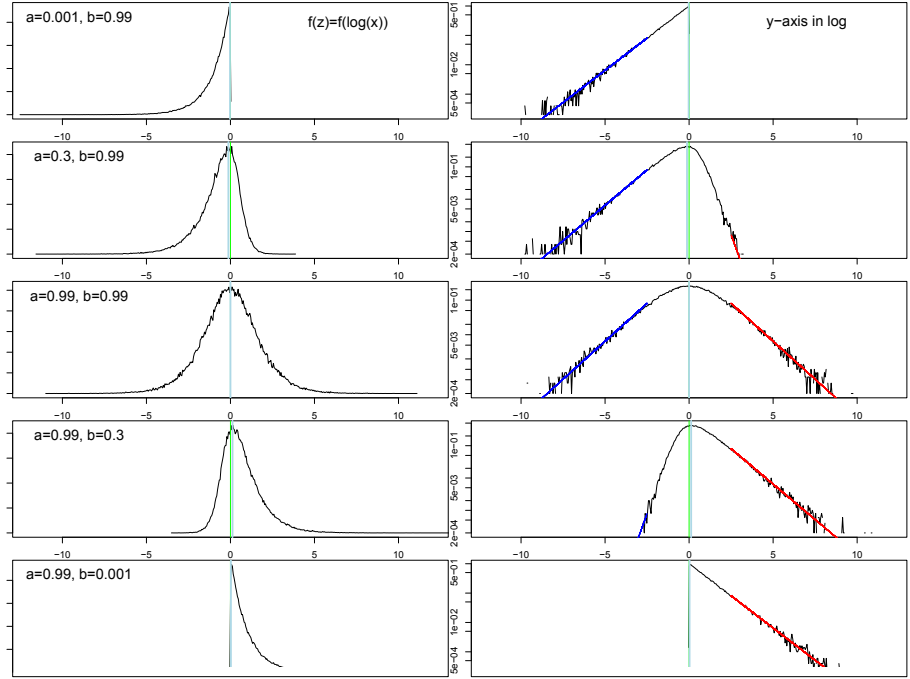}
\end{center}
\caption{
\label{fig3}
Histogram (empirical pdf) of sample points simulated from the BRF 
(with $A=1$) at various parameter values, with $z=\log(x)$ as the x-axis:
(A) a=0.001, b=0.99 (close to a uniform distribution);
(B) a=0.3, b=0.99;
(C) a=b=0.99 (Lavalette distribution);
(D) a=0.99, b=0.3;
(E) a=0.99, b=0.001 (close to a Pareto distribution).
The right panel corresponds to the left panel with $y$-axis in log.
The blue left fitting line is $e^{z/b}/b$ and red fitting line on the right for $e^{-z/a}/a$.
}
\end{figure}

Numerically, we can either approximate the distribution with a histogram of the sampled values, or by the numerical approximation of the functional form. Fig.\ref{fig3} shows histograms of (log-transformed) randomly sampled values according to the procedure described in section \ref{random-BRF} (with $A=1$, and various $a$, $b$ values).  When the histogram which approximates $f_Z(z)$ is log-transformed, on the right panel of Fig.\ref{fig3}, it is clear that the decay from the peak on both tails is exponential. We use the fittings  $e^{z/b}/b$ on left and $e^{-z/a}/a$ on the right and see that both fit the empirical $f_Z(z)$ perfectly. 

\begin{figure}[th]
\begin{center}
    \epsfig{file=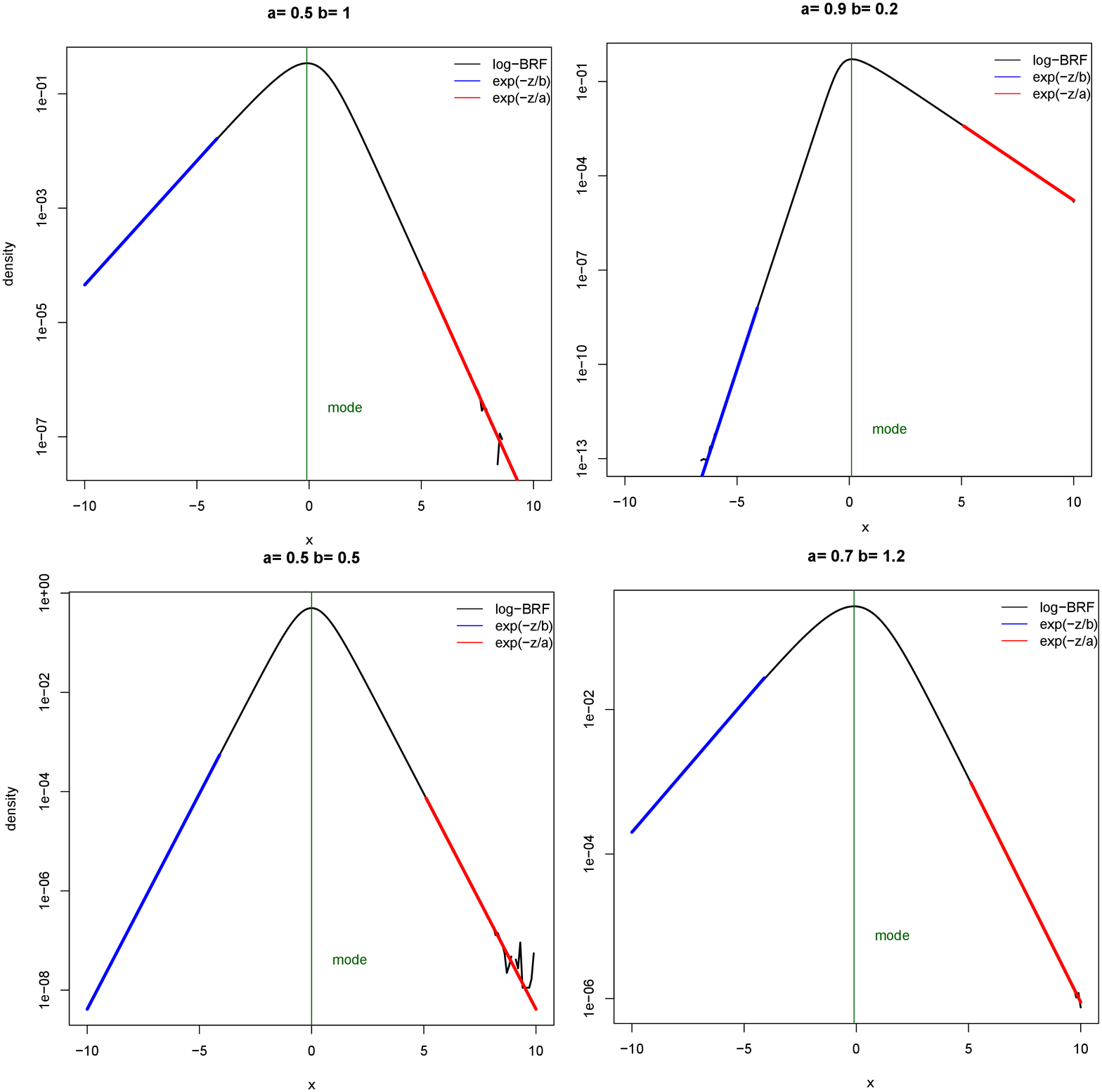, width=12cm}
\end{center}
\caption{
\label{fig4}
Numerical reconstructions of the $Z$ pdf for different parameter values. We also show $e^{z/b}/b$ (blue) and $e^{-z/a}/a$ (red), as well as the theoretical mode (green). The $y$ axis is on logarithmic scale. 
}
\end{figure}

As an independent check, we use the numerical approximation procedure described in section \ref{numerical-approx} to generate the $f_Z(z)$ at several parameter values and the result is shown in Fig.\ref{fig4}. Again, the blue line on the left side of the peak is $e^{z/b}/b$ and the red line on the right side of the peak is $e^{-z/a}/a$; both fit perfectly with the numerical approximation of the $f_Z(z)$.

Now we show analytically that both tails of $f_Z(z)$ are approximately exponential
without knowing the analytic form of $f_Z(z)$. First, we derive the relationship
between $z=\log(x)$ and $u$:
\begin{equation}
\label{eq-z-u}
z=\log(x)= \log \left(A \frac{(1-u)^b}{u^a} \right)= \log(A)+b\log(1-u)-a\log(u)
\end{equation}
The pdf $f_X(x)$ is the negative derivative of $u(x)$:
\begin{equation*}
f_X(x)= -\frac{d u(x)}{dx}= - \frac{ \frac{dz}{dx}}{ \frac{dz}{du}} = - \frac{\frac{1}{x}}{ -\frac{b}{1-u} - \frac{a}{u}}
\end{equation*}
Then (note that the cdf is $F= 1-u$):
\begin{equation}
\label{eq-fz-u}
f_Z(z)=xf_X(x)= \frac{1}{ \frac{b}{1-u}+\frac{a}{u}} = \frac{1}{ \frac{b}{F}+\frac{a}{1-F}}
\end{equation}
Near the right tail (from Eq.(\ref{eq-z-u}), at the $u \rightarrow 0 $ limit, $z \approx \log(A)-a\log(u)$):
\begin{equation}
\label{eq-right-side}
f_Z(z) \approx \frac{1-F}{a}= \frac{u}{a} \approx \frac{e^{ - \frac{z-\log(A)}{a}}}{a} = \frac{A^{1/a}}{a}e^{-\frac{z}{a}}
\quad \mbox{if} \quad z \gg  0.
\end{equation}
Near the left tail (similarly from Eq.(\ref{eq-z-u}), at the $u \rightarrow 1 $ limit, $z \approx \log(A) +b \log(1-u)$):
\begin{equation}
\label{eq-left-side}
f_Z(z) \approx \frac{F}{b}= \frac{1-u}{b} \approx \frac{e^{  \frac{z-\log(A)}{b}}}{b}
=\frac{A^{-1/b}}{b}e^{\frac{z}{b}}
 \quad \mbox{if} \quad z \ll  0.
\end{equation}
Now we not only know that $f_Z(z)$ decays exponentially on both tails (and log-exponentially, thus algebraically for $f_X(x)$), but also we know the parameters $b$ and $a$ control the left and right side of the exponential decay rate respectively. When $a=b$, the two decay rates are equal. When $a > b$, the decay on the right side of the peak decays slower and vice versa. This result is consistent with the probability partition by the peak discussed in the previous subsection: when $a > b$, the decay on the right side of the peak is slower, thus occupying more probability area underneath and the other way around.
The same result for  $f_X(x)$ is also obtained in \cite{hankin2006new}.


\subsection{Cubic term  and asymmetry near the peak of $f_Z(z)$ }

\indent

We have just shown that $f_Z(z)$ decays exponentially on both sides
of the peak with (usually) different decay rates. This leads to an asymmetry around the peak, which requires higher order terms beyond the quadratic term in a Taylor expansion. Our previous results on quadratic approximation of Lavalette function can not be applied.

Using Taylor expansion, we have 
\begin{equation}
\label{fZ-taylor-1}
 f_Z(u) = \frac{1}{(\sqrt{a}+\sqrt{b})^2} - \frac{1}{\sqrt{ab}} (u-u_0)^2 +
  \frac{b-a}{ab}(u-u_0)^3 +  O(u-u_0)^4,
\end{equation}
near the peak $u_0=\sqrt{a}/(\sqrt{a}+\sqrt{b})$.  From the relationship $z=\log A +b\log(1-u) -a\log u$, we also have that near $u_0$,
$$
z(u) \approx \log A -a\log u_0 +b \log (1-u_0) -(b+2\sqrt{ab}+a)(u-u_0),
$$
which we can solve for $u$,
$$
u \approx \frac{1}{(\sqrt{a}+\sqrt{b})^2} \left[ a+\sqrt{ab} +\log A +b\log(1-u_0)-a\log u_0\right] - \frac{1}{(\sqrt{a}+\sqrt{b})^2}z.
$$
Substituting this in Eq.(\ref{fZ-taylor-1}) and rearranging terms we get a polynomial expansion for $f_Z$ near the mode $z_0 = \log A + b\log (1-u_0) -a \log u_0$,
\begin{equation*}
\label{fZ-taylor}
f_Z(z) = c_1-c_2(z-z_0)^2+c_3(z-z_0)^3 + O(z-z_0)^4,
\end{equation*}
where
$$
\begin{array}{l}
\displaystyle c_1 = \frac{1}{(\sqrt{a}+\sqrt{b})^2},\\
\\
\displaystyle c_2 = \frac{1}{\sqrt{a}(\sqrt{a}+\sqrt{b})^4\sqrt{b}},\\
\\
\displaystyle c_3 = \frac{\sqrt{a}-\sqrt{b}}{a(\sqrt{a}+\sqrt{b})^5b}.\\
\end{array}
$$

\begin{figure}[th]
\begin{center}
    \epsfig{file=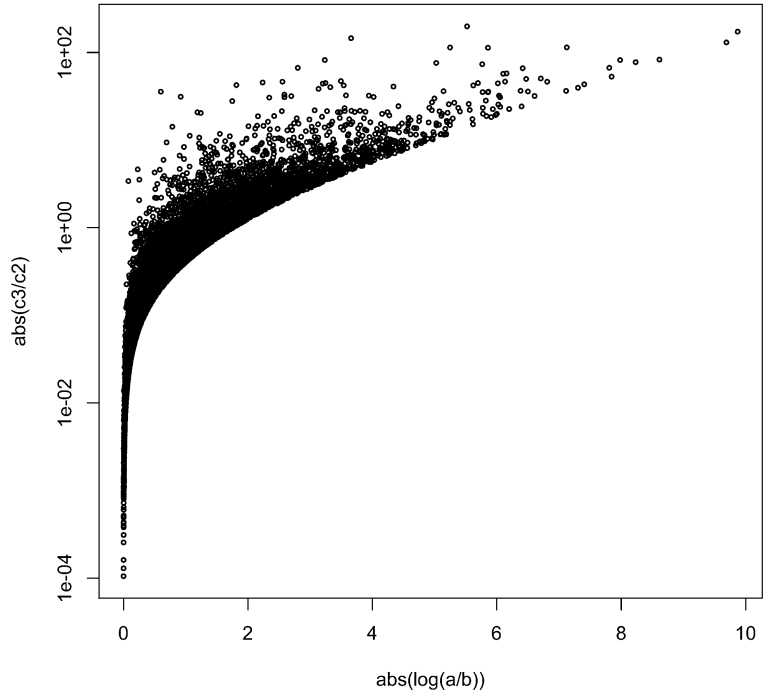, width=10cm}
\end{center}
\caption{
\label{fig5}
Ratio of cubic and quadratic coefficients for the Taylor expansion of $f_Z$ near the peak as a function of $\log(b/a)$. When $a$ and $b$ are different, the cubic terms dominates, producing an asymmetry near the peak of the density function. 
}
\end{figure}

Since the quadratic term is symmetric around the peak, whereas the cubic term is not, we expect $c_3/c_2$ will increase if $a$ and $b$ are more different. Fig.\ref{fig5} shows $c_3/c_2$ as a function of $|\log(b/a)|$. Indeed, the cubic term dominates the quadratic term when $a$ and $b$ are very different.

\section{The BRF family of distributions}

Once we have defined the log-BRF distribution, we can formally define that the continuous random variable $X$ follows a \emph{BRF distribution} with parameters $A > 0$ and $a,b \geq 0$ if the random variable $Z=\log X$ follows a log-BRF distribution. According to this definition, the pdf of $X$ must have non-negative  support (because $e^Z$ is always a non-negative number). \\

The pdf of $X$ can be written in terms of the Fox $H$ function, with the formula 

\begin{equation*}
f_X(x) = \displaystyle \Fox{1}{1}{1}{2}{\frac{x}{A}}{(-a,a/A)}{(1-b,b/A)\,(b-a-1),(a-b/A)}.
\end{equation*}

\noindent See Appendix \ref{Fox-section} for a detailed derivation of this expression and for the characteristic function of this distribution. This makes the BRF a particular case of the Fox-H distribution, defined as the family of non-negative distributions whose pdf can be expressed by a function of this class. The Fox-H family of distributions is closed under products and quotients, thus suggesting possible generating mechanisms of the BRF distribution \cite{carter1977distribution}.

\subsection{Moments of $f_X(x)$ are given by a Beta function}

\indent

Recall the special function called Euler integral (or Beta function):
\begin{equation*}
B(x,y) =\int_{0}^{1} t^{x-1} (1-t)^{y-1} dt.
\end{equation*}
The moments of $f_X(x)$ can be represented by this special function: 
\begin{eqnarray*}
E[X^n] & = & \displaystyle \int_0^\infty x^n f(x)dx 
 \nonumber \\
 & = & \displaystyle -\int_{u(0)}^{u(\infty)}\left[ A\frac{(1-u)^b}{u^a}\right]^n \frac{ d u(x) }{dx} dx
 \nonumber \\
 & = & \displaystyle A^n\int_{u=0}^{u=1}  u^{-na} (1-u)^{nb}  du \nonumber \\
 & = & A^n B( 1-na, 1+nb) = A^n \frac{\Gamma(1-na) \Gamma(1+nb)}{ \Gamma( nb-na+1)},
\end{eqnarray*}
which is finite if $n<\frac{1}{a}$.

\subsection{Median and peak position (mode) for $f_X(x)$}

\indent

The aim here is to find the solution for $d f_X(x)/ dx=0$. First, the first derivative of $f_X(x)$
is the second derivative of $u(x)$: $d f_X(x)/ dx= - d^2 u/dx^2$. Secondly, there is a relation
between the second derivative of a function and that of its inverse function:
\begin{equation*}
\frac{d^2 x(u)}{ du^2} = - \frac{ \frac{d^2 u}{dx^2}}{ \left( \frac{dx}{du}\right)^3}
\end{equation*}
so we now can solve $d^2 x(u)/du^2=0$ instead.
Take the second derivative of Eq.(\ref{BRF}), we have
\begin{equation*}
b(b-1) + 2ab \frac{(1-u)}{u} + a(a+1) \frac{(1-u)^2}{u^2} =0
\end{equation*}
whose solution is:
\begin{equation*}
u_0= \frac{a - \sqrt{ \frac{ab}{a-b+1}}}{ a-b  }
\hspace{0.3in}\mbox{ if $a> b-1$ and $a \ne b$}
\end{equation*}
An equivalent solution can be derived by using $v=1/u$:
\begin{equation*}
u_0= \frac{a (a+1) }{ a(a-b+1) + \sqrt{ ab(a-b+1)}  }
\end{equation*}
When $a=b$, setting the second derivative to zero is to solve a linear equation:
\begin{equation*}
(a-1) + (a+1) \frac{(1-u)}{u}  =0
\end{equation*}
which leads to the solution
\begin{equation*}
u_0= \frac{1+a}{2}, \hspace{0.2in}\mbox{or} \hspace{0.1in} F_0 = \frac{1-a}{2}
\hspace{0.3in}\mbox{ if $a=b$}
\end{equation*}

Inserting the $u_0$ value to Eq.(\ref{BRF}) we have the peak position in $f_X(x)$:
\begin{equation*}
x_0 =
\left\{
\begin{array}{ccc}
A (a-b)^{a-b}  \frac{\left( \sqrt{ \frac{ab}{a-b+1}} -b \right)^b}{ \left( a-\sqrt{\frac{ab}{a-b+1}} \right)^a}  
& \mbox{if} & a > b-1  \hspace{0.1in} \mbox{ and } a \ne b \\
 & & \\
A \left( \frac{1-a}{1+a} \right)^a
& \mbox{if} & a=b  \\
\end{array}
\right.
\end{equation*}
Comparing to the similar result on peak position of $f_Z(z)$ in Eq.(\ref{eq-peak-z}), we see that $z_0$ is not equal to $\log(x_0)$.

\subsection{Randomly sampling variables from the BRF distribution}
\label{random-BRF}
\indent

Because $1-u(x)$ is the cdf, randomly sampling $u$ values from a (0,1) uniform distribution
also randomly samples cdf values. Since $x$ is a simple function of $u$, the corresponding
$x$ values thus obtained represent a random sampling from the BRF distribution. The general idea on this procedure can be found in \cite{rvg}. A simple R ({\sl https://www.r-project.org/})  code can be (e.g.) the following:
\begin{verbatim}
	a <- 0.5
	b <- 1.2
	A <- 1
	nump <- 1E6
	u <- runit(nump)
	x <- A*(1-u)^b/u^a
\end{verbatim}

\subsection{Numerical approximation of the pdf}
\label{numerical-approx}

Getting the pdf of the BRF distribution requires inverting the rank-size function $x(u)$, which can be seen as a root-finding step, then differentiating the function $u(x)$ . The first step is a problem we cannot analytically solve, but there are several algorithms to do it numerically. Thus, we can combine a numeric root-finding procedure with some numerical differentiation rule to establish an algorithm to numerically approximate the pdf of a BRF. For the sake of illustration, we propose the following algorithm: 
(i) Choose a tolerance interval $t$ for root finding and a step size $h$ for numerical differentiation;
(ii) use a root-finding method to solve Eq.(\ref{BRF}) for $u$, i.e.,
 find the root of $\left(\frac{x}{A}\right)u^a-(1-u)^b=0$, then $F(x)=1-u(x)$;
(iii) use finite-difference coefficients to compute $f(x)= \frac{dF(x)}{dx}$.

\begin{figure}[th]
\begin{center}
    \epsfig{file=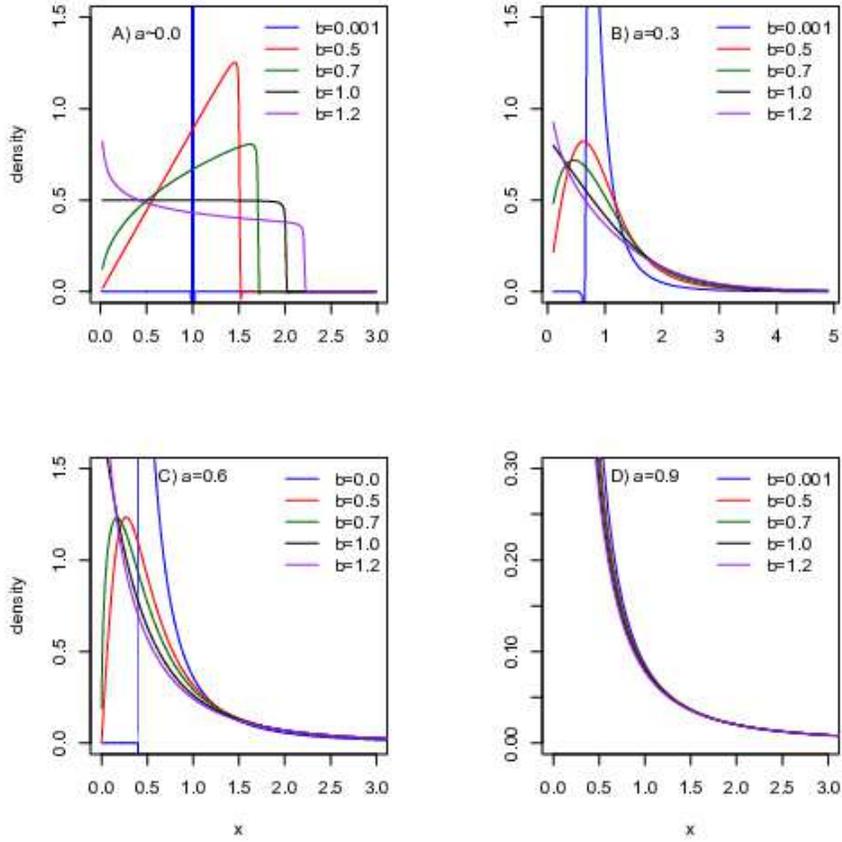, width=12cm}
\end{center}
\vspace{-1.0cm}
\caption{
\label{figS1}
Numerical approximations to the BRF pdf by the bisection plus finite difference coefficients algorithm.
}
\end{figure}

There are several combinations of root finding (step (ii)) and numerical differentiation techniques (step-(iii)) that can be used. Consider, for instance, bisection method for step (ii), which guarantees convergence, and one-dimensional five-point stencil for step (iii).  The error of this algorithm  with these two numerical methods are discussed below; from one side, we have the finite difference approximation for the derivative,
$$
f'(x) =\frac{ L_h[F(x)]}{12h} + O(h^4),
$$
where $L_h[F(x)] = -F(x+2h)+8F(x+h)-8F(x-h)+F(x-2h)$. Let $F_a(x)$ be the numerically determined cdf from step (ii), implying that this differs from the real value of $F(x)$ in less than the tolerance interval, $|F_a(x)-F(x)|\leq t$. Consequently, $|L[F_a(x)]-L[F(x)]|\leq 18t$, and we can compute the total error:
$$
\begin{array}{lll}
\left| f(x)-L[F_a(x)]/12h \right| & = & \left| f(x)-L[F_a(x)]/12h +L[F(x)]/12h -L[F(x)]/12h \right|\\
  & \le & \left| f(x) - \frac{L[F(x)]}{12h} \right|  + \left| \frac{L[F_a(x)]}{12h} - \frac{L[F(x)]}{12h} \right|\\
 & \le & O(h^4) + \frac{3t}{2h},\\
\end{array}
$$
meaning that the error of the algorithm is of order $\mbox{max}(h^4,\frac{t}{h})$ We show some examples of this procedure in Fig.\ref{figS1}.

Notice in Fig.\ref{figS1}A that, when $a=0$, the pdfs are defined over a finite interval, as expected  from section \ref{BRF-uniform}. We can see the constant random variable with its delta pdf (blue line), as well as the uniform distribution (black line). At $b=1$, the pdfs shift from being increasing to decreasing. The pdfs are always unimodal,  but for $a \in (0,1)$ the peak can be either at $x=0$ or to the right, depending on the value of  $b$, as can be seen in Fig.\ref{figS1}B and C. When $b=1$, the location of the maximum shifts  from a positive to a zero value.  Notice also that, when $b\approx 0$, the pdf exhibits a typical  power-law behavior with a cut-off, as expected from section \ref{BRF-Pareto}. Interestingly, the parameter $a$  completely dominates the behavior of the pdf when it is close to $1$: as can be seen in Fig.\ref{figS1}D, pdfs with different values for $b$ become hard to distinguish from one another.


\section{Using the properties of the BRF and log-BRF distributions in data analysis}

\subsection{Modeling log-returns of financial indexes with log-BRF distribution}

Understanding the properties of $f_Z(z)$ provides us with very simple approaches to distinguish the distribution type in real data. In order to analyze data that may be well described by a log-BRF distribution (unimodal, support on the entire real axis, exponential decay on both tails) we propose the following pipeline: (1) plot the histogram of a set of observations $\{ z_i \}$ with the $y$-axis in log scale; (2) examine the shape of the histogram with $y$ in log scale to infer the type of the distribution that produced the observations.

As an illustration of this pipeline, consider the analysis of logarithmic returns of financial assets or indexes. Recall that if $S_t$ denotes the price of a certain financial asset at time $t$, then $S_t/S_{t-1}$ is called the gross one-period simple return and $\log S_t/S_{t-1}$ is called the logarithmic gross simple return of the asset (which we simply call the log-return of the asset). Classical models such as the Black-Scholes-Merton model imply that log-returns are normally distributed; however, several observations point to the presence of a fatter than normal tail on these distributions (\cite{stanley2000introduction,sornette2017stock}). This seems like a good candidate to test our suggested pipeline.

We analyzed daily log-returns over a 30 year period (from April 24 1989 to April 18 2019) of four different financial indexes: Dow Jones Industrial Average, NASDAQ Composite, Standard and Poor 500 and Russell 2000. Data were downloaded from \url{https://finance.yahoo.com/}.

We propose two different methods to estimate the parameters $a$ and $b$ of the BRF-density from the data (we do not know the exact form of the pdf, so more common methods, such as maximum likelihood, are not available). First we utilize the method of moments, by equating the first two sample central moments to the known expressions of $E[Z]$ and $E[Z^2]$. The estimator we get from this method are
$$
\begin{array}{c}
\hat{a} = \displaystyle \frac{\bar{Z}}{2} + \frac{[\bar{Z}^2(\pi^2-12)+12S^2]^{\frac{1}{2}}}{2\pi},\\
 \\
\hat{b} = \displaystyle -\frac{\bar{Z}}{2} +
\frac{[\bar{Z}^2(\pi^2-12)+12S^2]^{\frac{1}{2}}}{2\pi},
\end{array}
$$
\noindent where $\bar{Z}$ and $S^2$ are sample mean and variance respectively. In order to reduce the bias of these estimators, we utilize the Jackknife re-sampling technique, by aggregating the estimates of reduced samples.\\

The second method we propose considers the fact that, in semi-log scale, a log-BRF decays linearly on both tales, with gradient $+1/b$ on the left tail and $-1/a$ on the right. Thus, we can restrict to one of these two domains, approximate the density $f$ from the histogram and and fit the linear models $\log f \sim \frac{1}{b} z$ and  $\log f \sim -\frac{1}{a} z$. \\

\begin{figure}[th]
\begin{center}
    \epsfig{file=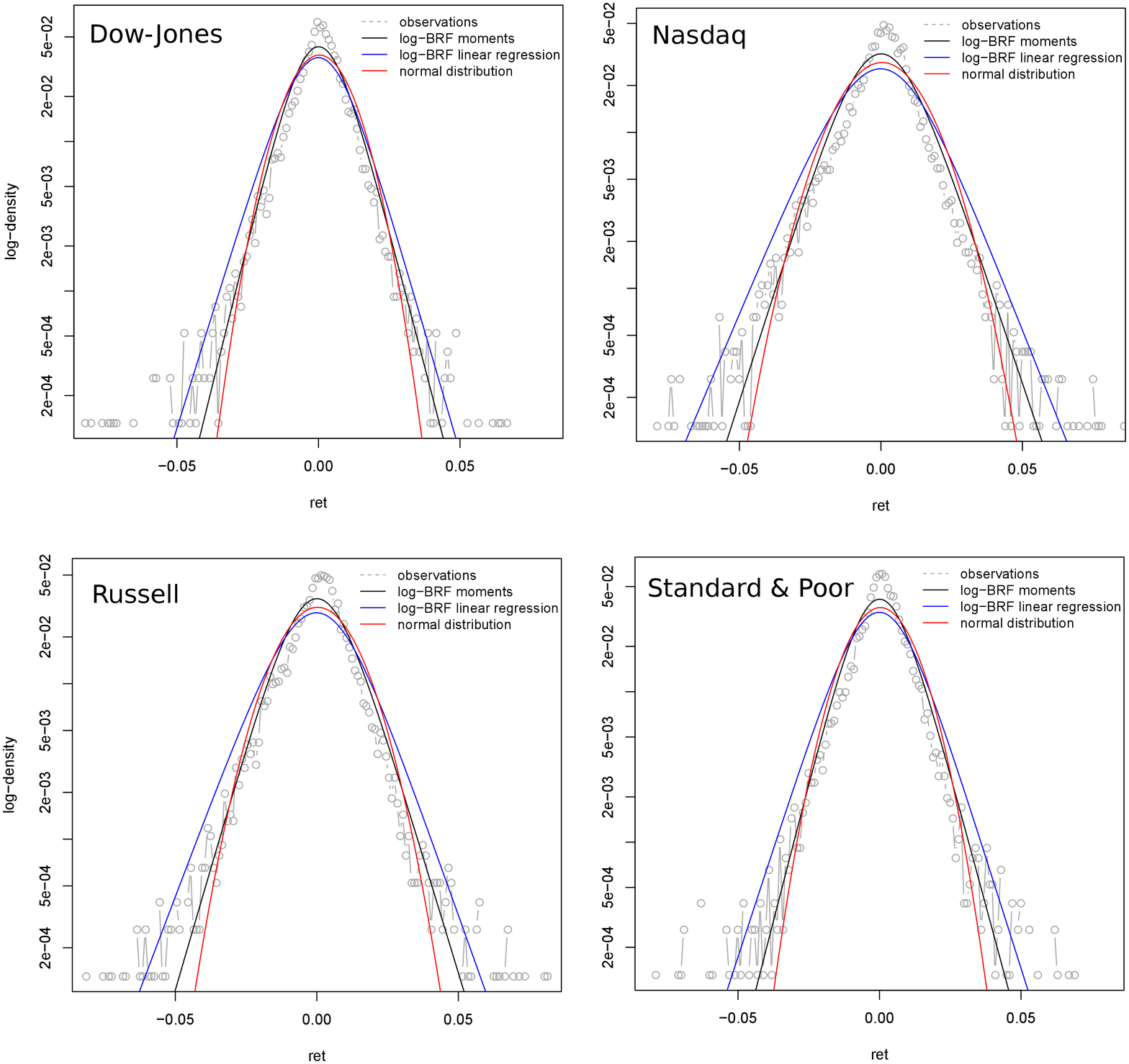, width=12cm}
\end{center}
\caption{ Histograms and fitted log-BRF and normal distributions for 30 years daily
log-returns of four different financial indexes. The $y$ axis is in logarithmic scale.
\label{returns}
}
\end{figure}

We show in Fig.\ref{returns} the results of these analysis. For each of the four financial index that we modeled, we show in semi-log scale the histogram of the observations (gray), the fitted log-BRF distribution with parameters estimated by the moments method (black) and the linear regression technique (blue); we also show the fitted normal distribution (red). In all four cases we observe a similar behavior: neither log-BRF nor the normal/log-normal fits the data satisfactorily at the peak of the distribution, but the normal/lognormal model tends to sub-estimate the tails, which log-BRF does not. We also note that, apparently, the linear regression technique yields better estimations than the Jackknife-moments method; this is no surprise, since for the former we are censoring the data, restring the observations to a different domain for each parameter. 

\subsection{Modeling urban population with a BRF distribution}

There are several examples in the literature that utilize the BRF distribution (or DGBD rank-size function) to fit data that looks like a power-law, but with a break at the upper tail ( high rank / small size regime). In addition to these numerous examples, we propose here the following pipeline, that takes advantage  of the properties of the log-BRF distribution: (1) log-transform the data $\{ x_i \}$ to $\{ z_i =\log(x_i) \}$; (2) plot the histogram of $\{ z_i \}$ with the $y$-axis in log scale; (3) examine the shape of histogram with $y$ in log scale to infer the type of the distribution $\{ x_i \}$ may follow.

For example, if the histogram with $y$ in log scale falls from the peak only on the right side as a straight line, the distribution should be approximately a one-sided power-law. If the histogram falls from both sides with roughly equal slope, the distribution is approximately lognormal, or lognormal-like. If the histogram falls from both sides linearly with different slopes, the distribution is a BRF ($a \ne b$). If there are not much data on one sides of the peak, we do not have enough evidence to claim the distribution to be a BRF.

\begin{figure}[th]
\begin{center}
    \epsfig{file=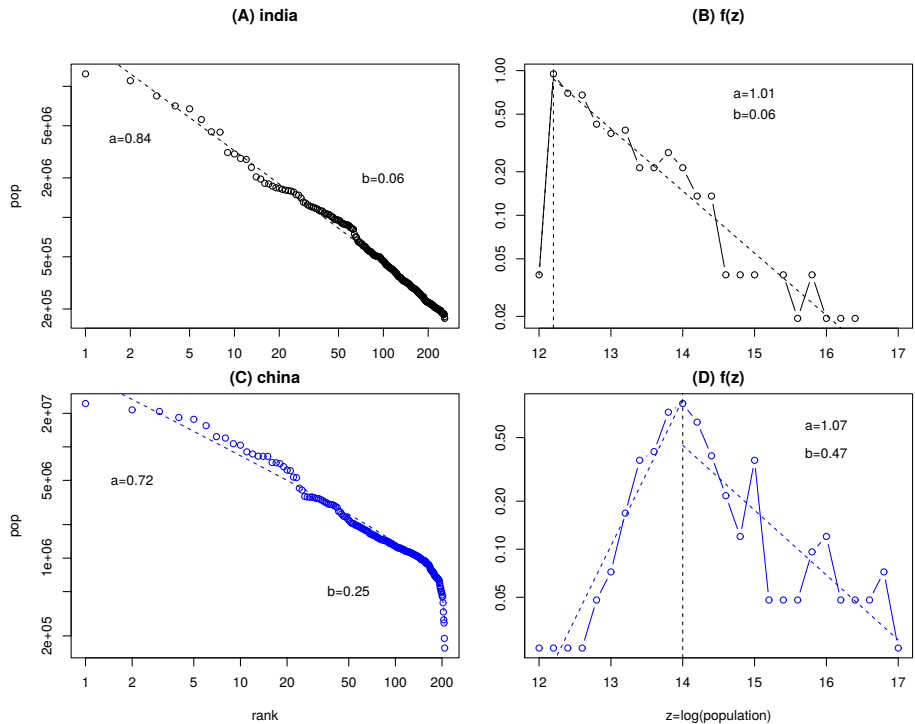, width=12cm}
\end{center}
\caption{Rank-size plot of India (A) and China (C) city population and fitted BRF; histogram of log-population for India (B) and China (D) and fitted BRF. The $a$ and $b$ parameters are estimated through different approaches in each representation. 
\label{fig6}
}
\end{figure}

Fig.\ref{fig6} shows the India city population (2011 census) and China urban population (2010 census) represented in two different plots: the rank-population plot (Fig.\ref{fig6}A,C), and the histogram of log-population (Fig.\ref{fig6}A,C). The BRF in the rank-population plot can be fitted directly by the linear regression $\log(x) = C -a  \log(r) + b\log(r_2)$ where $x$ is the population
data, $r$ is the rank ($r=1$ for the largest city), and  $r_2 \equiv  N+1-r$.

In the histogram of $\log(x)$, Fig.\ref{fig6}B,D already provides some essential information without fitting. For example, Fig.\ref{fig6}B indicates that India city population is mostly a one-sided power-law, as the peak is close to the left and the fall off from the peak on the left is very steep (indicating a small $b$ value). On the other hand, China urban population follows BRF better as the peak is closer to the middle, and the decay rate from both sides are different. Fitting the two sides by exponential decay leads to a qualitatively similar estimation of $a,b$ as the direct fitting in rank-population plot. The differences of the fitting $a,b$ values between the two approaches may be due to several factors: the choice of bin size in the histogram, the skipping of zero frequency in the histogram (as the logarithmic value diverges), etc.

Note that the left end of Fig.\ref{fig6}A,C becomes right tail in Fig.\ref{fig6}B,D, and vice versa.  Fig.\ref{fig6}B,D) also shows that the most likely India city population is 0.2 millions, and the most likely China urban population is 1.2 millions, in this particular data set.

\section{Discussion and Conclusions}

\indent

The Discrete Generalized Beta Distribution (DGBD) \cite{mansilla,gustavo}, is
developed mainly in the camp of Zipf's law study \cite{pedro-arxiv,plos-lav}, which  focus on one-sided power-law. Little attention was paid to the fact that, since DGBD is actually a two-sided power law, the better representation should be the pdf showing the peak and fall-off from the peak on both sides. We show in this paper that the crucial step in doing so is to log-transform the measured variable $x$, $z=\log(x)$, then the shape of the pdf becomes more clear. 

By recognizing the peak in the middle of the distribution, we also recognize a fact that BRF, unlike the Pareto distribution, does not diverge at the peak, which 
is a problem for many other usage of one-sided power law functions. That divergence  in one-sided power law is usually dealt with by adding a cutoff (truncation) \cite{clauset}.

The two approximate expressions of the log-BRF pdf on two sides of the peak (Eq.(\ref{eq-right-side}) and
Eq.(\ref{eq-left-side})) can be rewritten as $(x/A)^{-1/a}/a$ (for $x >  A$) and $ (x/A)^{1/b}/b$
(for $x < A$), where $x=e^z$. Using $f(z) dz = f(x)dx$ and $dz/dx=1/x$ (for simplicity, let us assume $A=1$), 
the BRF pdf in $x$ space is approximately $f(x) \approx \left( \frac{1}{x} \right)^{1+1/a}/a$ (for $x >1$),
and $\left( \frac{1}{x} \right)^{1-1/b}/b$ (for $x < 1$). It illustrates again the
importance of using the log-transformed variable, as the left size of the ``peak" may not
decay in $f(x)$ if $b > 1$.

The functional form above appears in the literature recurrently with different names: log-Laplace distribution \cite{uppuluri,kotz},  skewed log-Laplace distribution \cite{kozu}, and double Pareto distribution \cite{reed01,athari}. These are exponential functions for
log-transformed variables with a sharp angle at the peak. In 2004, Reed and Jorgensen proposed a new distribution which combines the double Pareto and the normal distribution, called ``double-Pareto-lognormal" (DPLN) distribution  \cite{reed04}, which has seen many applications \cite{reed02,csuros,giesen,haj,toda17} . This function, like BRF,
has a smooth transition between the two sides at the peak. However, DPLN has four fitting parameters, 
two for the double Pareto and two for the normal distribution,
whereas the BRF distribution only has two, without the need to use another normal distribution
to smooth the function. The lesser number of parameters also makes BRF a flexible function to be utilized in data analysis,
as we demonstrated in this paper through several examples.

With a better understanding of the pdf of BRF, we now can clarify confusions with the 
Beta distribution whose pdf is $\propto x^{a-1} (1-x)^{b-1}$ \cite{jambunathan}. The range of $x$ is (0, $\infty$) 
for BRF, but is (0,1) in Beta distribution.  BRF is unimodal, whereas Beta distribution
may have two peaks (or two singularity points). When $a=b$ both BRF and Beta distribution
are symmetric. However, BRF falls from the peak as inverse power-law, whereas Beta distribution
as a concave quadratic function or its power. Needless to say, BRF and Beta distribution are not
the same. 

It is interesting that DGBD or BRF has been independently discovered multiple times
in other fields. We are aware of at least two publications, one by Gilchrist 
in section 1.6 of \cite{gilchrist2000statistical}, and another by Hankin and Lee (but attributed to  
Davies through private communication) in \cite{hankin2006new}, in the context of quantile functions \cite{parzen}. 
In a quantile function, the quantile value of a random variable $X$ is expressed as a function of the cumulative probability $0 \le p \le 1$. 
This framework to represent a probability distribution is exactly in parallel to the rank-size plot,
one of the two versions used to illustrate Zipf's law \cite{urzua,wli-zipf,rousseau-zipf}, if we convert
$p$ to the normalized rank $u=1-p$. Neither \cite{gilchrist2000statistical} nor \cite{hankin2006new}
explored the possibility of using the log-transformed variable $Z=\log X$ and its impact on
functional form of the probability distribution. This is however the trick that
allowed us to derive many analytic results not available in
\cite{gilchrist2000statistical,hankin2006new}.

BRF is an example of a distribution where the closed-form analytic expression of the pdf or cdf is not generally available
(i.e., it is not expressed by a finite number of ``well-known" functions). However, its pdf is expressible
as a function of its cdf or rank variable \cite{hankin2006new}. Here we further show that BRF's pdf can be expressed in terms of
Fox-H functions, which provides a rigorous definition of both functions (with and without log transformation
of the $x$ variable). This task is accomplished by using the characteristic functions. 

In conclusion, we provide a most comprehensive analysis of the continuous-rank version of 
DGBD or BRF, pointing out the key step in log-transforming the $X$ variable. We have obtained expressions for the
mean, median, mode, variance and other quantities of the log-BRF. We have established the basic shape
of the distribution. The parallelism between BRF and double-Pareto distribution, skewed log-Laplace distribution,
and double-Pareto-lognormal distribution makes it one more useful function to fit real-world data with an
appropriate statistical feature. 


\section{Acknowledgments}

This work was partially supported by UNAM-PAPIIT grant IN108318. OF is a grant holder of the UNAM-DGAPA Postdoctoral Scholarships Program at CEIICH, UNAM. WL acknowledges the support from Robert S Boas Center for Genomics and Human Genetics.


\bibliographystyle{ieeetr}

\bibliography{references-arxiv}


\section{Appendixes}

\subsection{Appendix A - Analytic expression of $f_X(x)$ when $a=2b$ and $b=2a$}

\indent

In the next few appendixes, we present a few analytic results without using the 
independent variable log-transformation.  As will be seen, the derivation of these 
analytic formula is more tedious. The first example is the analytic expression of
$f_X(x)$ when $a$ is exactly twice the value of $b$.
The steps towards an analytic solution is: (i) obtain the inverse function
$u(x)$ from $x(u)$ by solving an algebraic equation; (ii) since cdf is $1-u(x)$,
$$
x = A \left( \frac{1-u}{u^2} \right)^b
$$
or
$$
x^{1/b} u^2 +u -1 =0
$$
The solution is:
\begin{equation*}
u(x) = \frac{\sqrt{1+4x^{1/b}}-1}{2x^{1/b}}
\end{equation*}
Take the derivative:
\begin{equation*}
f_{a=2b}(x) = -\frac{du(x)}{dx} = 
\frac{ \sqrt{1 + 4x^{1/b} }-1 }{2bx \cdot x^{1/b}} - \frac{1}{bx \sqrt{1 + 4x^{1/b} } }
\end{equation*}

The second example is when $b=2a$:
$$
x = A \left( \frac{F^2}{1-F} \right)^a
$$
or
$$
x^{-1/a} F^2+ F - 1 =0 
$$
with solution
\begin{equation*}
F(x)  = \frac{x^{1/a} \left(\sqrt{1+4x^{-1/a}} -1 \right)}{2}
\end{equation*}
Take the derivative:
\begin{equation*}
f_{b=2a}(x)= \frac{dF(x)}{dx}= \frac{x^{1/a} \left(\sqrt{1+4x^{-1/a}} -1 \right) }{2ax} 
- \frac{ 1  }{ ax(\sqrt{1+4x^{-1/a}} -1)}
\end{equation*}


\subsection{Appendix B - The probability density function $f_X(x)$ when $a=3b$ and $b=3a$}

When $a=3b$, we have a relationship between $x$ and $1-F=u$ ($F$ is the cumulative density function):
\begin{equation}
\label{eq-a=3b}
x = A \left( \frac{1-u}{u^3} \right)^b,
\hspace{0.1in} \mbox{with $x >0$, $A >0$, $u \in (0,1)$}
\end{equation}
or
\begin{equation*}
u^3 + \left( \frac{x}{A} \right)^{-1/b}u = \left( \frac{x}{A} \right)^{-1/b}
\end{equation*}
The left-hand-side term is negative ($-1$) when $u=0$, and positive ( $(x/A)^{1/b}$) when $u=1$.
Also, the first derivative (slope),  $ 3 (x/A)^{1/b} u^2 +1$, is always positive. Therefore, Eq.(\ref{eq-a=3b}) has a single
solution for $x$ when $u \in (0,1)$. Using Tartaglia's trick \cite{katscher}
(for $u^3+pu =q$ with $p=q$), the solution is:
\begin{equation*}
u = \sqrt[3]{ \sqrt{ \left(\frac{C}{2} \right)^2+ \left(\frac{C}{3} \right)^3 } + \frac{C}{2}}
 - \sqrt[3]{ \sqrt{ \left(\frac{C}{2} \right)^2+ \left(\frac{C}{3} \right)^3 } - \frac{C}{2}}
\end{equation*}
where $C \equiv (x/A)^{-1/b}$.

The pdf is the derivative $dF/dx=-du(x)/dx$ (note $dC(x)/dx= C/(bx)$):
\begin{equation*}
f_X(x)= 
\frac{1}{6bx}
\left[\frac{C+ \frac{C^2/2 +C^3/9}{ \sqrt{ (C/2)^2+(C/3)^3}} }{ (\sqrt{ (C/2)^2 +(C/3)^3} +C/2)^{2/3}  }
 + \frac{C- \frac{C^2/2 +C^3/9}{ \sqrt{ (C/2)^2+(C/3)^3}} }{ (\sqrt{(C/2)^2 +(C/3)^3} - C/2)^{2/3}  }
\right] 
\end{equation*}

When $b=3a$, the procedure is very similar:
\begin{equation*}
\label{eq-b=3a}
x = A \left( \frac{F^3}{1-F} \right)^a,
\hspace{0.1in} \mbox{with $x >0$, $A >0$, $F \in (0,1)$}
\end{equation*}
or
\begin{equation*}
F^3 + \left( \frac{x}{A} \right)^{1/a}F = \left( \frac{x}{A} \right)^{1/a}
\end{equation*}
The solution is
\begin{equation*}
F = \sqrt[3]{ \sqrt{ \left(\frac{D}{2} \right)^2+ \left(\frac{D}{3} \right)^3 } + \frac{D}{2}}
 - \sqrt[3]{ \sqrt{ \left(\frac{D}{2} \right)^2+ \left(\frac{D}{3} \right)^3 } - \frac{D}{2}}
\end{equation*}
where $D \equiv (x/A)^{1/a}$.

Take the derivative :
\begin{equation*}
f_X(x)= \frac{dF}{dx}=
\frac{1}{6ax}
\left[
\frac{ 
 D+ \frac{ D^2/2 +D^3/9}{ \sqrt{ (D/2)^2+ (D/3)^3     } } 
}
{ (\sqrt{ (D/2)^2 +(D/3)^3} +D/2)^{2/3}}
+
\frac{
 D- \frac{ D^2/2 +D^3/9}{ \sqrt{ (D/2)^2+ (D/3)^3     } } 
}{ (\sqrt{ (D/2)^2 +(D/3)^3} +D/2)^{2/3}}
\right]
\end{equation*}

Exact pdf expressions for $a=4b$ and $b=4a$ are still possible to derive with this methodology, but they are extremely cumbersome to be included here.

\subsection{Appendix C - Characteristic function and a closed-form representation of $X$ and $Z$}
\label{Fox-section}

\noindent The pdf of Z is the inverse Fourier transform of the characteristic function. Hence we can get an integral formula for the pdf of Z,
$$
\displaystyle f_Z(z) = \frac{1}{2\pi}\int_{-\infty}^{\infty}e^{-izt}\psi_Z(t)dt = 
\frac{1}{2\pi}\int_{-\infty}^{\infty}e^{-izt}A^{it} B(1-iat,1+ibt).
$$

\noindent With this formula and the relationship $f_X(x)dx=f_Z(z)dz$ we can write an integral formula for the pdf of $X$,
$$
\begin{array}{lll}
f_X(x) & = & \frac{1}{x}f_Z(z(x))\\
& & \\
 & = & \displaystyle \frac{1}{2\pi x}\int_{-\infty}^{\infty} e^{-it\log x} A^{it} B(1-iat,1+ibt)dt\\
& &  \\
& = & \displaystyle \frac{1}{2\pi x}\int_{-\infty}^{\infty} x^{-it}A^{it} B(1-iat,1+ibt)dt \\
& & \\
& = & \displaystyle \frac{1}{2\pi x}\int_{-\infty}^{\infty} \left( \frac{A}{x} \right)^{it} B(1-iat,1+ibt)dt.\\
\end{array}
$$

\noindent We are taking again the principal branch of the logarithm. Recall the relationship between the Beta and the Gamma function,
\begin{equation}
\label{beta-gamma}
B(x,y)=\frac{\Gamma (x) \Gamma (y)}{\Gamma (x+y)}.
\end{equation}

\noindent Now, recall the definition of the Fox H-function. The H-function is defined by the Mellin-Barnes type contour integral
$$
\displaystyle \Fox{m}{n}{p}{q}{x}{(a_1,A_1),...,(a_p,A_p)}{(b_1,B_1),...,(b_q,B_q)} 
= \frac{1}{2\pi i}\int_C h^{m\,n}_{p\,q}(s)x^sds,
$$
\noindent where the function $h$ is defined by
\begin{equation*}
\label{small-h}
h^{m\,n}_{p\,q}(s) = \frac{\displaystyle \prod_{j=1}^m \Gamma (b_j-sBj) \prod_{j=1}^n \Gamma(1-a_j+sA_j)}
{\displaystyle \prod_{j=m+1}^q \Gamma(1-b_j+sB_j) \prod_{j=n+1}^p \Gamma (a_j-sA_j)}.
\end{equation*}
\noindent An empty product is considered as unit. Here, $x\neq 0$, $x^s= exp (s(\log |x| +i \,arg(x))$; $p$, $m$, $n$ and $q$ are integers such that $0 \leq m\leq q$ and $0\leq n\leq p$; $A_j$ and $B_j$ are positive real numbers: $a_j$ and $b_j$ are real or complex numbers such that no pole of $\Gamma (b_j-sB_j)$ coincides with a pole of $\Gamma(1-a_j-sA_j)$. Finally, $C$ is a contour on the complex plane running from $c-i\infty$ to $c+i\infty$ for  some real constant $c$ such that the poles from $\Gamma (b_j-sB_j)$ lie to left of $C$ and all poles of $\Gamma(1-a_j-sA_j)$ lie to the right of $C$. Conditions for the existence of the contour $C$ and for the existence and analyticity of the function $H$ are comprehensively enumerated in \cite{mathai}. \\

\noindent The following properties of the H-function can be deduced directly from the definition:
\begin{enumerate}

\item
$
\Fox{m}{n}{p}{q}{x^\mu}{((a_\lambda,A_\lambda))}{((b_\lambda,B_\lambda))} = 
\frac{1}{\mu}\Fox{m}{n}{p}{q}{x}{((a_\lambda,\frac{A_\lambda}{\mu}))}{((b_\lambda,\frac{B_\lambda}{\mu}))}
\quad \mbox{for} \quad \mu>0.
$

\item
$
\displaystyle \Fox{m}{n}{p}{q}{\frac{1}{x}}{((a_\lambda,A_\lambda))}{((b_\lambda,B_\lambda))} = 
\Fox{n}{m}{q}{p}{x}{((1-b_\lambda,B_\lambda))}{((1-a_\lambda,A_\lambda))}
$

\item
$
x^\sigma \Fox{m}{n}{p}{q}{x}{((a_\lambda,A_\lambda))}{((b_\lambda,B_\lambda))} = 
\Fox{m}{n}{p}{q}{x}{((a_\lambda+\sigma A_\lambda, A_\lambda))}{((b_\lambda+\sigma B_\lambda,B_\lambda))}
\quad \mbox{for} \quad \sigma \in \mathbb{C}.
$
\end{enumerate}

\noindent By using (\ref{beta-gamma}) and making the substitution $s=it$, we can write the pdf of $X$ as
$$
f_X(x) = \displaystyle \frac{1}{x} \frac{1}{2\pi i} \int_{-\infty}^{\infty} \left( \frac{A}{x} \right)^{s} \frac{\Gamma (1-as) \Gamma (1+bs)}{\Gamma(2+bs-as)}ds.
$$

\noindent Notice that we can write 
$$
\frac{\Gamma (1-as) \Gamma (1+bs)}{\Gamma(2+bs-as)} = \frac{\Gamma (b_1-sB_1) \Gamma (1-a_1+sA_1)}{\Gamma(a_2-SA_2)}
$$
\noindent by doing $b_1=1$, $B_1=a$, $a_1=0$, $A_1=b$, $a_2=2$ and $A_2=a-b$. Therefore, we can write the pdf of the BRF random variable in terms of an $H$ function,

\begin{equation*}
f_X(x) = \frac{1}{x} \Fox{1}{1}{2}{1}{\frac{A}{x}}{(0,b)\,(2,a-b)}{(1,a)},
\end{equation*}

\noindent for $x>0$.  It is possible to write this pdf in another form by means of the properties of the $H$ function,
\begin{equation*}
\begin{array}{lll}
f_X(x) & = & \displaystyle \frac{1}{A} \left( \frac{A}{x} \right) \Fox{1}{1}{2}{1}{\frac{A}{x}}{(0,b)\,(2,a-b)}{(1,a)}\\
 & & \\
 & = & \displaystyle \frac{1}{A} \Fox{1}{1}{2}{1}{\frac{A}{x}}{(b,b)\,(2+a-b,a-b)}{(a+1,a)} \\
 & & \\
 & = & \displaystyle \frac{1}{A} \Fox{1}{1}{1}{2}{\frac{x}{A}}{(-a,a)}{(1-b,b)\,(b-a-1,a-b)}\\
 & & \\
 & = & \displaystyle \Fox{1}{1}{1}{2}{\frac{x}{A}}{(-a,a/A)}{(1-b,b/A)\,(b-a-1),(a-b/A)}.
\end{array}
\end{equation*}

\noindent Finally, we can also write the pdf of $Z=\log X$ in terms of the $H$ function,

\begin{equation*}
f_Z(Z) = \displaystyle \Fox{1}{1}{1}{2}{\frac{e^z}{A}}{(0,a)}{(1,b)\,(-1,a-b)}.
\end{equation*}

\end{document}